\documentclass[prb,twocolumn,showpacs,preprintnumbers,amsmath,amssymb]{revtex4} 

\usepackage{amsmath,amssymb,amsfonts}
\usepackage{bm}
\usepackage{graphicx}
\usepackage{sidecap}
\usepackage{times}
\usepackage{gensymb}
\usepackage{color}

\usepackage[colorlinks,bookmarks=false,citecolor=blue,linkcolor=red,urlcolor=blue]{hyperref}

\begin{document}
\title{Correlation effects in transport properties of interacting nanostructures}
\author{A.~Valli$^1$, G.~Sangiovanni$^{1,2}$, A.~Toschi$^1$, and K.~Held$^1$.}
\affiliation{$^1$ Institute of Solid State Physics, Vienna University of Technology, 1040 Vienna, Austria\\
$^2$ Institute for Theoretical Physics and Astrophysics, University of W\"urzburg, Am Hubland, D-97074 W\"urzburg, Germany}
\date{\today}

\begin{abstract}
We discuss how to apply many-body methods to correlated nanoscopic systems, 
and provide general criteria of validity for a treatment at the dynamical mean field theory (DMFT) approximation level, 
in which local correlations are taken into account, while non-local ones are neglected. 
In this respect, we consider one of the most difficult cases for DMFT, namely for a quasi-one-dimensional molecule such as a benzene ring. 
The comparison against a numerically exact solution shows that non-local spatial correlations are relevant only 
in the limit of weak coupling between the molecule and the metallic leads and of low inter-atomic connectivity, 
otherwise DMFT provides a quantitative description of the system. 
As an application we investigate the role of correlations on electronic transport in quantum junctions, 
and we show that a local Mott-Hubbard crossover is a robust phenomenon in sharp nanoscopic contacts. 
\end{abstract}
\maketitle

\section{Introduction}
Strong electronic correlations are a big challenge in condensed matter theory. 
In the case of bulk materials, dynamical mean field theory (DMFT) \cite{DMFT} 
turned out to be a big breakthrough - at least in three dimensions.
The reason for this is that local, time-dependent electronic correlations are taken into account accurately. 
This local part is the major contribution of electronic correlations at least for high enough coordination or dimensions or at elevated temperatures. 
For realistic materials calculations on the other hand, DMFT has been merged with density functional theory 
in the local density approximation (LDA). \cite{LDADMFT} 
All these calculations are done in the thermodynamic limit, i.e., for an infinitely extended crystal. 

In nanoscopic systems the confinement of electrons into low-dimensional structures 
is expected to enhance correlation effects compared to bulk materials. 
Since complex nanoscopic systems are nowadays experimentally available, 
a microscopic modeling of them together with a reliable solution method 
taking electronic correlations into account is highly desirable. 
In the context of transport through nanoscopic systems also the detailed modeling of the reservoirs 
is required to accurately describe the experimental data. 

Very small nanoscopic systems and molecules consisting only of a few atoms can still be calculated exactly, 
i.e. the low energy effective Hamiltonian for these can be solved, or for even simpler di- or tri-atomic molecules 
numerically exact solutions are possible e.g. by quantum Monte Carlo (QMC) \cite{hirschPRL56,maierRMP77,gullRMP83} 
or configuration interaction. 
However, as soon as the nanoscopic or molecular systems become somewhat more complex this is 
not possible any longer and, hence, a reliable approximation is needed. 
Recently, there have been efforts to apply DMFT also to finite systems 
such as nanoscopic structures and molecules connected to reservoirs. \cite{valliPRL104,linPRL106} 
Since nanoscopic systems are very different from the bulk, at this early stage, we have first of all to learn how reliable DMFT 
for describing electronic correlations in nanoscopic and molecular systems is.

A general scheme for treating correlated nanoscopic systems should also include non-local spatial correlations beyond DMFT. 
In this respect cluster \cite{DCA} and diagrammatic extensions \cite{toschiPRB75,heldPTPS176,kataninPRB80,toschiAP523,others} have been developed. 
Among the diagrammatic extensions, the dynamical vertex approximation (D$\Gamma$A) \cite{toschiPRB75,heldPTPS176,kataninPRB80,toschiAP523} 
represents a systematic improvement beyond DMFT, as it allows to calculate the non-local part of the self-energy 
under the assumption of locality for the 2-particle fully irreducible vertex. 
Recently it has been shown that the fully irreducible vertex is computationally accessible \cite{rohringer1202.2796}, 
e.g. via the numerical solution of an Anderson impurity model (AIM). 
However a full calculation for nanostructures at the D$\Gamma$A approximation level, 
including the solution of the parquet equations \cite{toschiPRB75,heldPTPS176,kataninPRB80,toschiAP523,parquet}, is indeed computationally expensive.

In this paper we will focus on the DMFT approximation level, that can be seen as a special case of a general scheme 
that we call ``nano-D$\Gamma$A'', as described in Ref. \onlinecite{valliPRL104}. 
In Section \ref{Sec:method}, we outline the method and discuss the connection with related or alternative approaches. 
In Section \ref{Sec:benzene}, in order to understand the reliability of DMFT for nanoscopic systems, we compare it extensively to a numerically exact solution, 
in an interesting case of a quasi one-dimensional molecule (benzene ring), and provide general criteria of validity for the approximation. 
In Section \ref{Sec:QPC} we show the potentiality of the method applying it to single atom quantum junctions, 
namely to quantum point contact (QPC) of different sizes.
Finally, Section \ref{Sec:conclusion} provides a summary and outlook.

\section{Method}\label{Sec:method}
As pointed out in the introduction, we are interested in a nanoscopic system consisting of sites (e.g.\ atoms) $i$ with 
an inter-site hybridization (hopping) $t_{ij}$, a local Coulomb repulsion $U_{i}$ and (optionally) a coupling $V_{i\nu k}$ 
to some non-interacting environment, describing metallic leads contacted to the nanostructure. 
The Hamiltonian hence reads
\begin{eqnarray}
H&=& \sum_{ij\sigma} t_{ij} c^{\dagger}_{i\sigma} c^{\phantom{\dagger}}_{j\sigma}
+ \sum_i U_{i}  c^{\dagger}_{i\uparrow} c^{\phantom{\dagger}}_{i\uparrow} c^{\dagger}_{i\downarrow} c^{\phantom{\dagger}}_{i\downarrow} \nonumber \\ &&+
 \sum_{i \nu k \sigma} V_{i  \nu k}  c^{\dagger}_{i\sigma}  l^{\phantom{\dagger}}_{\nu k\sigma} + h.c. + 
 \sum_{\nu k \sigma} \epsilon_{\nu k}  l^{\dagger}_{\nu k \sigma}  l^{\phantom{\dagger}}_{\nu k\sigma},
\label{eq:Hamiltonian}
\end{eqnarray}
where $c^{\dagger}_{i\sigma}$ ($c^{\phantom{\dagger}}_{i\sigma}$) and $l^{\dagger}_{\nu k \sigma}$ ($l^{\phantom{\dagger}}_{\nu k \sigma}$)
denote the creation (annihilation) operators for an electron with spin $\sigma$ on site $i$ 
and in lead $\nu$ state $k$ with energy $\epsilon_{\nu k}$. 
For the purposes of the present paper the one-band Hamiltonian (\ref{eq:Hamiltonian}) is enough, because the main goal is to validate our approach.
The extension to multi-orbital problems is straightforward and below we will highlight the corresponding modifications 
to the general scheme. 

Let us just recall here, that a numerically exact solution of this problem suffers from a non-polynomial growth of the computational effort 
with the system size and is hence limited by severe restrictions on the number of sites. 
Therefore, if one aims at dealing with complex structures made of more than a few coupled sites, some kind of approximation is needed.
A DMFT-like approach with a suitable approximation may be able to deal with a large number of coupled correlated sites, but 
in order to apply DMFT -and its extensions- to nanoscopic systems one needs to define a proper local impurity problem, 
whose solution (usually numerical) represents the bottleneck of the algorithm.

What one can do is to reduce the $N$-impurity Anderson problem (\ref{eq:Hamiltonian}) onto a 
set of independent auxiliary AIMs, one for each of the $N_{\rm ineq}\!\le\!N$ inequivalent atoms in the nanostructure. 
Each AIM is a local problem that can be numerically solved to compute the $N_{\textrm{ineq}}$ correspondent (local) self-energies, 
the knowledge of which allows to build a DMFT self-energy for the nanosctructure. The process is embedded into a self-consistency loop. 
This way the overall computational effort is heavily reduced, depending only linearly on $N_{\rm ineq}$, 
that in case of highly symmetric structures may be much lower than $N$.
This procedure has however the drawback that non-local correlations within the nanostructure, 
which are expected to be relevant in low dimensions, are neglected. 
The treatment -when necessary- requires a more sophisticated analysis beyond the DMFT level. 

The starting point of the method is the Green's function of the whole nanostructure (including the leads) 
$\hat{G}(z)$, which is a matrix in the site space, $z$ being a (complex) variable indicating the (Matsubara) frequency.
The generic matrix element of its inverse reads
\begin{equation} \label{Dyson}
 \displaystyle{ \big\{\hat{G}^{-1}\big\}_{ij}(z) = z\delta_{ij} - t_{ij} - \sum_{\nu k} \frac{V^{\phantom{*}}_{i \nu k}V^{*}_{j \nu k}}{z - \epsilon_{\nu k}} - \Sigma_{ij}(z) }, 
\end{equation}
where $\hat{\Sigma}(z)$ is the self-energy matrix describing the interaction between the impurity electrons.
In the multi-orbital case the matrices $\hat{t}$, $\hat{V}$, $\hat{\Sigma}$ and $\hat{G}$ would also depend on orbital indices.
All information about the geometry of the nanoscopic system is included in the hopping and in the hybridization matrices. 
At the model level it is therefore straightforward to implement even extremely complex nanostructures. 
The input may come as well from an {\em ab-intio} calculation, 
e.g. a local density approximation (LDA) projected to Wannier orbitals, \cite{kunesCPC181}  
allowing for realistic calculations of nanoscopic systems and a quantitative comparison with experiments.

The general flowchart of the method is shown in Fig.\ref{fig:flowchart} and is described below in more detail:

\begin{figure}[ht]
\begin{center}
\includegraphics[angle=0, width=0.9\columnwidth]{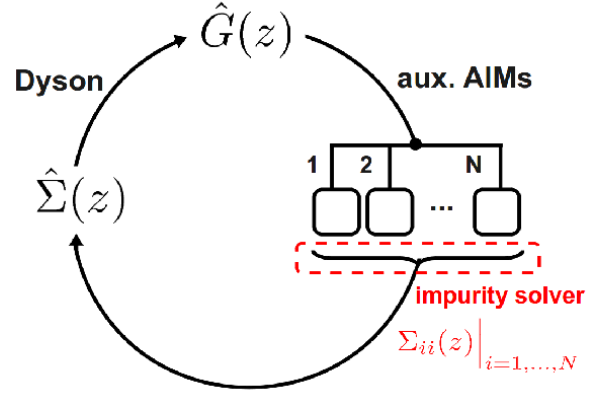}
\caption{Flowchart of the DMFT self-consistency schemes for nanoscopic systems. 
The approximation consists in mapping the (non-local) problem of the whole nanostructure into a set of independent AIMs. 
The solution of these local problems yields a set of local self-energy, which can be used to define a DMFT self-energy for the nanostructure.}
\label{fig:flowchart}
\end{center}
\end{figure}

(i) the first step consists of the definition of a local problem for each of the inequivalent atoms of the nanostructure, 
by means of the relation
\begin{equation}\label{eq:weissfield}
 {\cal G}^{-1}_{0 i}(z) = {\Big[ \big\{\hat{G} \big\}_{ii}(z) \Big]}^{-1} + \Sigma_{ii}(z).
\end{equation}
The dynamic Weiss field ${\cal G}_{0 i}(z)$, $i\!=\!1, ..., N_{\rm ineq}$, is built inverting the $i$-th block of $\hat{G}$ 
and it contains the information of the environment of site $i$, i.e. the rest of the nanostructure.
In the multi-orbital case Eq. (\ref{eq:weissfield}) becomes obviously a matrix equation with orbital indices. %in the correlated-band subspace.

(ii) the numerical solution of each AIM yields a local (DMFT) self-energy $\Sigma_{ii}(z)$. 
All the $N_{\rm ineq}$ inequivalent self-energies are then collected and assigned to the corresponding equivalent sites as well, 
in order to build a self-energy matrix which is diagonal in the site index
\begin{equation}
 \hat{\Sigma}(z)\!=\!{\rm diag} \big( \Sigma_{11}(z),\Sigma_{22}(z), ... \ , \Sigma_{NN}(z) \big).
\end{equation}
The self-energy $\hat{\Sigma}(z)$ is then plugged into Eq. (\ref{Dyson}) in order to compute the Green's function of the whole nanostructure  
and the process is iterated self-consistently till convergence.

The approximation involved in the present scheme is already known in the literature, 
and similar schemes have been applied to different kind of systems. 
An approach for quasi one-dimensional systems is the chain-DMFT by Biermann {\it et al.}, \cite{biermannPRL87}
where a system of weakly coupled (equivalent) chains is replaced by a single effective chain, coupled to a self-consistent bath. 
More in general, the idea is suitable to the study of inhomogeneous systems, 
and has been applied to, e.g., the study of bulk materials in the presence of two-dimensional interfaces by Potthoff and Nolting \cite{potthoffPRB60}, 
as well as to the case of LDA+DMFT calculations with locally-inequivalent atoms within the unit cell (see e.g. \onlinecite{dasPRL107}). 
Another noticeable case is its application to ultracold atoms on optical lattices, using the so-called real-space DMFT (R-DMFT), 
by Snoek {\it et al.}, \cite{snoekNJP10} where the inhomogeneity comes from the external, spatially dependent, trapping potential, 
applied to an otherwise translationally invariant lattice. 
The present approach is similar to the R-DMFT, the difference being that in our case each site is also coupled to a non-interacting bath, 
and a possible inhomogeneity arises not due to an external potential, 
but from the geometry or even the chemical composition of the nanostructure itself.

The application of DMFT to nanoscopic systems, on the other hand, has been already attempted 
following alternative ways. A nano-DMFT scheme has been already proposed by Florens \cite{Florens07}, 
relying however on a specific cayley-tree geometry. 
Realistic calculations of strongly correlated transition metal nanoscopic devices and of correlated ad-atoms on surfaces 
have been also recently carried out by Jacob {\it et al.} \cite{jacobDMFT} and by Surer {\it et al.} \cite{surerPRB85}, respectively.

\section{Results}
In the following we will apply the presented DMFT method to various nanoscopic systems, 
in order to test the reliability of the approximation and to explore its potentialities. 

In this spirit, we extend the results presented in our previous work \cite{valliPRL104} 
computing several physical quantities, such as the occupations, local and non-local self energies, 
in the case of a benzene ring, where we can compare DMFT to a numerically exact solution. 
This allows us to show that our method is highly reliable in a wide range of parameters, 
but also to shed light on the physical role of the (missing) non-local correlations, which are responsible for the breakdown of the approximation.

Thereafter, the most natural step is to show the suitability of the method for more complex nanosystems. 
We focus our attention on quantum junctions, in which electronic correlations are expected to be of importance 
due to both the confinement of electrons and the lack of a proper metallic screening at the atomic size contact. 
%Besides, we also study transport through a graphene nanostructure, a situation where we find correlation effects
%to be less relevant, since this nanostructure remains gapped even upon the inclusion of the leads and electronic correlations.
Even though in the literature many theoretical attempts to investigate transport at the mesoscopic scale can be found, 
the electron-electron interaction is usually either completely neglected or taken into account 
within possibly too simple approximation schemes. \cite{cuevasPRL80, scheerNat394, haefnerPRB77, wierzbowskaPRB72}

All the results presented below, both for DMFT and for the exact solution, were obtained using Hirsch-Fye QMC \cite{hirschPRL56} as an 
impurity solver for the AIM, 
unless otherwise stated. The use of a Hirsch-Fye algorithm limits us to high temperatures (about room temperature in the calculations presented here) 
but on the other hand it allows a quantitative comparison between the DMFT results and exact solution.

\subsection{Benzene-like ring}\label{Sec:benzene}
\begin{figure}[ht]
\begin{center}
\includegraphics[angle=0, width=0.60\columnwidth]{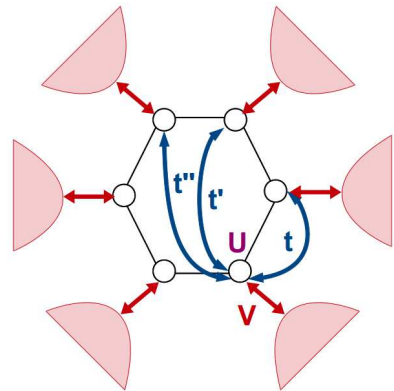}
\caption{(Color online) Scheme of the benzene ring. Empty circles represent correlated sites, 
with an on-site Hubbard repulsion $U$, 
connected between them via nearest neighbor ($t$) and longer range ($t'$, $t''$) tunneling channels, 
and to metallic leads via hybridization channels ($V$).}
\label{fig:toy_benzene}
\end{center}
\end{figure}

A rather standard system in which the problem of understanding quantum transport phenomena is addressed is
a benzene molecule, contacted with metallic (e.g. Au or Pt) electrodes.
Therefore we study a nanostructure made of six correlated sites in the geometry of a benzene-like one-dimensional ring, 
given by  Hamiltonian (\ref{eq:Hamiltonian}) with $i,j\!=\!1, ...,N\!=\!6$ and periodic boundary conditions. 
We consider here specifically a single orbital which may not be a bad approximation for the benzene $p_z$ orbital. \cite{footnote1} 
Our primary intention however, is to systematically test our approximation for a simple model rather than a realistic calculation.
Each site $i$ has a hybridization channel $V_{i \nu k}$ to a metallic lead (labeled $\nu$).  
In a typical experiment two sites of the benzene molecule might be contacted by metallic wires.
However, for the sake of simplicity, and to deal with a system where all sites are fully equivalent, 
we consider each site to be contacted in an equivalent way to its own lead, i.e., $V_{\imath \nu k}\!=\!V \delta_{i \nu}$.
%This local coupling in the site and lead space $V_{i \nu k}=V_{k} \delta_{i \nu}$, 
%corresponds to a hybridization in reciprocal space $\tilde{V}_{k}(q)$, which is the same for every $q$. 
The latter is not the only possibility to achieve the equivalence of all sites, but it represents a suitable configuration where one can 
study quantum electronic transport through a correlated nanostructure. 
%Another possibility to achieve the equivalence of all sites would be the case of the benzene molecule lying on a surface, 
%in which all sites are connected to the same metallic lead, 
%corresponding in reciprocal space to a unique non-zero coupling $\tilde{V}_{k}(q\!=\!0)$. 
A scheme of the nanostructure considered here is shown in Fig. \ref{fig:toy_benzene}. 

In the calculations two topologies of the hopping parameters are considered: (i) nearest neighbor hopping $t$ only (NN~t) and,
for studying the effect of a higher connectivity (number of neighbors) (ii) equivalent hopping
amplitude to all sites, i.e., nearest, next-nearest and next-next-nearest neighbor hopping $t\!=\!t'\!=\!t''$ (all~t). 
Of course the latter is a rather unrealistic configuration since the hopping amplitude in a real
molecule will decrease with distance, but it provides interesting insight 
into the validity of the approximation without introducing too many different hopping parameters.

\begin{figure}[t!]
\begin{center}
\includegraphics[angle=0, width=1.00\columnwidth]{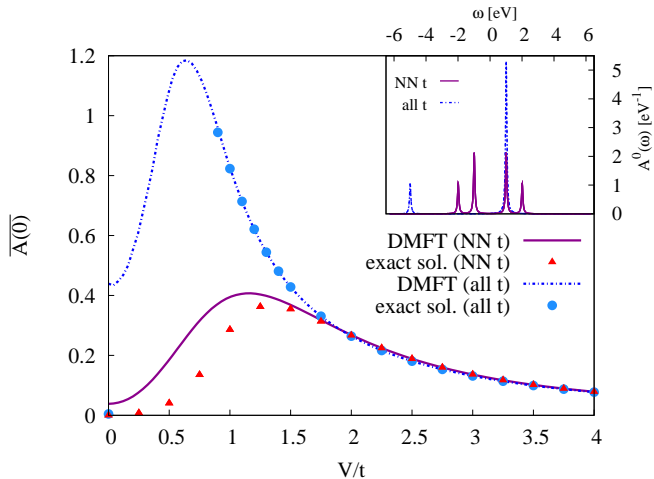}
\caption{(Color online) On-site spectral function $\overline{A(0)}$ as a function of $V/t$, 
comparing DMFT (lines) with the exact QMC solution (symbols) 
for both hopping to two nearest neighbors only (NN~t) and to all neighbors (all~t) configurations, 
at $U\!=\!5t$ and $T\!=\!0.05t$, taken from Ref. \onlinecite{valliPRL104}. 
Data for $V/t\!=\!0.0$ in the all~t configuration are obtained by an exact orthogonalization of the Hamiltonian.
Inset: non-interacting density of states $A^{0}(\omega)\!=\!-(1/\pi){\rm Im}G^0(\omega\!+\!\imath 0^+)$ 
for the isolated benzene molecule and both hopping configurations.}
\label{fig:A0}
\end{center}
\end{figure}

\begin{figure}[t!]
\begin{center}
\includegraphics[angle=0, width=1.00\columnwidth]{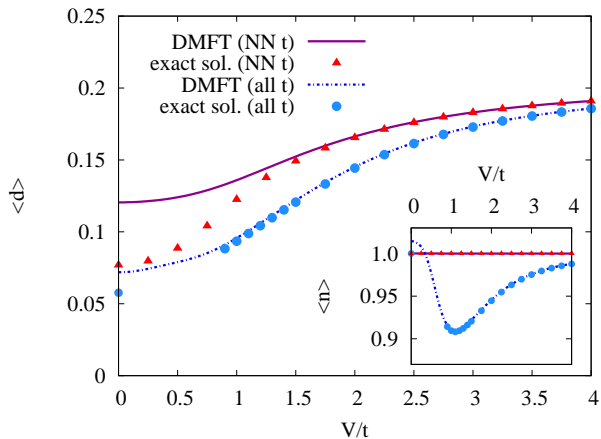}
\caption{(Color online) Double occupancies $\langle d \rangle\!=\!\langle n_{\uparrow}n_{\downarrow} \rangle$ as a function of $V/t$ 
at $U\!=\!5t$ and $T\!=\!0.05t$, comparing DMFT (lines) and the exact solution QMC (symbols) for both hopping configurations. 
Data for $V/t\!=\!0.0$ in the all~t configuration are obtained by an exact diagonalization of the Hamiltonian.
Inset: corresponding densities $\langle n \rangle\!=\!\langle n_{\uparrow}\!+\!n_{\downarrow} \rangle$.}
\label{fig:besetzung}
\end{center}
\end{figure}

We performed our DMFT calculations at fixed chemical potential, 
i.e., without considering the dependence on an applied gate voltage, 
assuming for the leads a flat density of states $\rho\!=\!1/2D$, where the half-bandwidth $D\!=\!2t$. 
We compute site-dependent densities, double occupations, and the on-site spectral function
\begin{equation}\label{eq:A0}
 \overline{A(0)}\!=\!\int{\rm d} \omega A(\omega) \cosh^{-1}{(\omega/2T)} = -\beta G(\beta/2), 
\end{equation}
which can be extracted directly by the QMC.
%, via the Fourier-transform of the spectral representation of the Green function. 
 
In the inset of Fig. \ref{fig:A0} we show non-interacting density of states $A^0(\omega)$ for the isolated molecule ($V/t\!=\!0$). 
In the NN~t case, the benzene ring is half-filled and insulating, 
the spectral function is symmetric with respect to the Fermi level and the gap given by the bonding and anti-bonding combination 
of the kinetic term in the Hamiltonian. The all~t case is also insulating but is not particle-hole symmetric. 
The results for $\overline{A(0)}$ as a function of the ratio between 
the hybridization strength $V$ and the absolute value of the hopping amplitude $t$ are shown in the main panel of Fig. \ref{fig:A0}. 
Looking at the NN~t topology, we observe that the agreement between the exact solution 
and DMFT is very good when the hybridization $V$ is large. 
In the limit $V\!\rightarrow\!\infty$, each atom forms a bound state with its own lead, 
hence the inter-site (non-local) correlations become essentially negligible and DMFT works well. 
The opposite molecular limit $V/t\!\equiv\!0.0$ is clearly the most difficult for DMFT. 
Indeed, the spectrum $\overline{A(0)}$ in Fig. \ref{fig:A0} differs from the exact solution, which is gapped, 
while DMFT shows a small finite spectral weight. 

One expects that, upon increasing the connectivity, the DMFT description improves. 
In order to check this, we consider the all~t topology as well. 
Indeed we see in Fig. \ref{fig:A0} that as the number of neighbors per atom is increased, 
no substantial difference to the exact solution can be found even in the intermediate region $V\!\sim\!t$
where deviations from the exact solution in the NN~t case are already visible. 
Note that below $V\!=\!0.8t$, the exact QMC solution is not available any longer due to the fermionic sign-problem,
therefore we cannot check the molecular limit for the all~t topology at this value of $U$ with such an impurity solver. 
The Hamiltonian of the isolated molecule can nevertheless still be diagonalized exactly also in the interacting case. 
The diagonalization predicts an insulating state also for the all~t configuration, meaning that the approximation, 
even in the high connectivity case, will break down at low enough hybridization. 

Similar agreement between DMFT and the exact solution 
is found for the site occupation, as shown in the inset of Fig. \ref{fig:besetzung}. 
In the NN~t topology, the effect of the interaction and of the hybridization on the spectrum is to redistribute the spectral weight, 
with respect to the non-interacting case, in such a way that the system stays half-filled. 
However when the band structure is changed, and other hopping channels beyond the NN one are included, the density becomes 
$t$, $U$, and $V$ dependent and the system may move away from half-filling. 
On the other side, concerning the double occupancy shown in the main panel, we can see that, for high connectivity and high values of $V/t$, 
corrections beyond DMFT are not important, while approaching the molecular limit 
the system is more spatially correlated than what DMFT suggests, overestimating double occupations.

\begin{figure*}[t!]
\begin{center}
\includegraphics[angle=0, width=2.00\columnwidth]{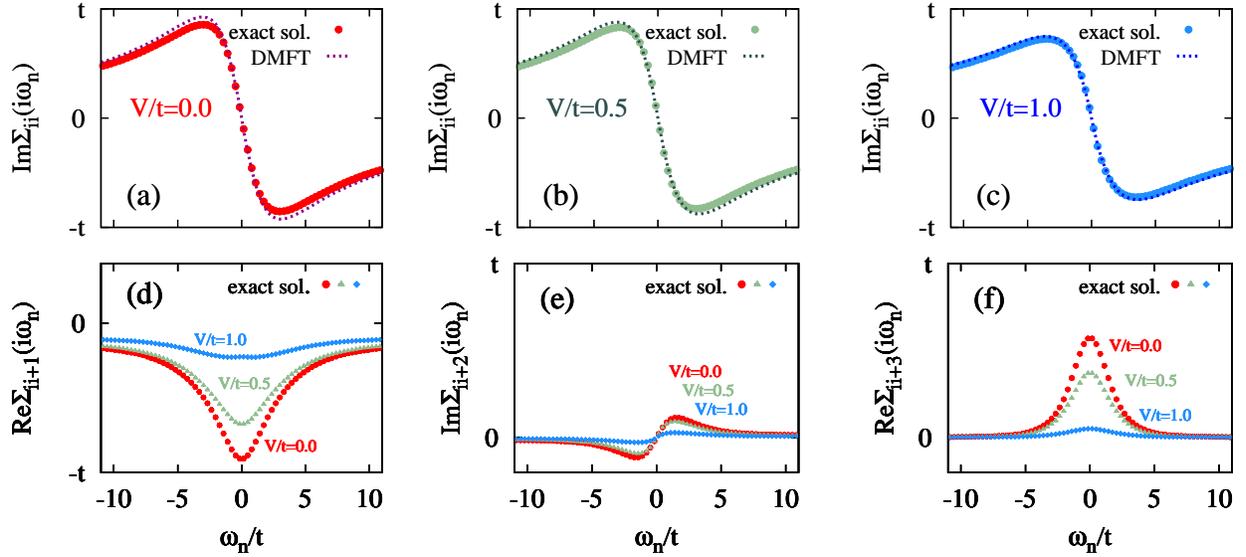}
\caption{(Color online) Self-energy in Matsubara representation for the benzene ring in the NN~t topology at $U\!=\!5 t$ and $T\!=\!0.05t$. 
In panels (a), (b), and (c) we show the evolution with $V/t$ of the imaginary part of the local self-energy, 
comparing DMFT (dashed lines) with the exact solution (symbols). 
Note that the real part is always identically zero due to the particle-hole symmetry at half-filling. 
In panels (d), (e), and (f) we show the non-local self-energy of the exact solution 
for nearest-neighbors ($i$,$i+1$), next-nearest neighbors ($i$,$i+2$), and next-next-nearest neighbors ($i$,$i+3$), respectively. 
All other non-local contributions are either identical to the ones shown here (since all sites are equivalent) or zero by symmetry. 
Note that $\Sigma_{i\neq j}$ is identically zero in DMFT.}
\label{fig:siw_nnt}
\end{center}
\end{figure*}

\begin{SCfigure*}
\centering
\includegraphics[angle=0, width=1.33\columnwidth]{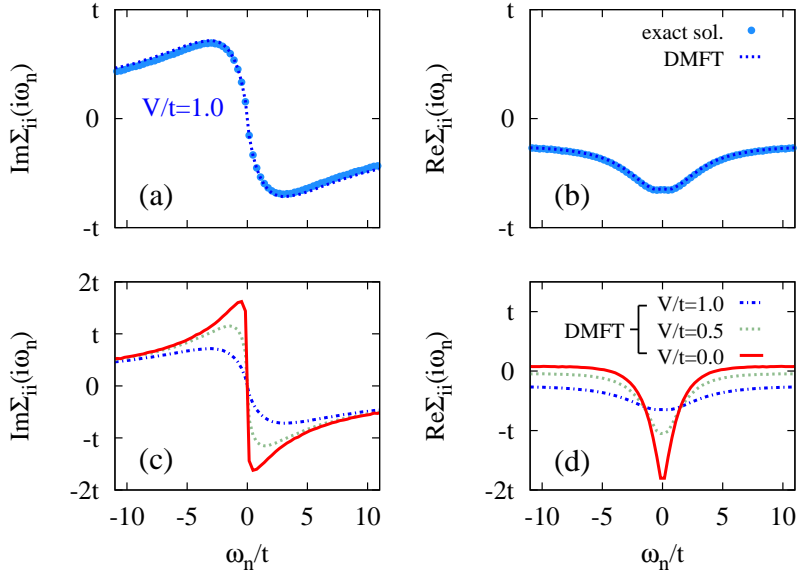}
\sidecaptionvpos{siedecap}{ht}
\caption{(Color online) Local self-energy in Matsubara representation for the benzene ring in the all~t topology at $U\!=\!5.0 t$ and $T\!=\!0.05t$. 
Upper panels: comparison between DMFT (lines) and exact solution (symbols) at $V/t\!=\!1.0$. 
Non-local contributions to the self-energy are negligible with respect to the local ones (almost two orders of magnitude smaller) 
in contrast to the NN~t case at the same value of $V$, and are therefore not shown.
Lower panels: evolution of the DMFT self-energy, comparing the curves of the upper panel 
to the ones obtained for lower values of $V/t$, in a region where no exact QMC solution is available.}
\label{fig:siw_allt}
\end{SCfigure*}

A very clear explanation of the overall agreement shown above, between DMFT and the exact solution, 
is provided by the comparison of the respective self-energies for both hopping topologies, 
shown in Fig. \ref{fig:siw_nnt} and \ref{fig:siw_allt}. 
As usual let us begin discussing the NN~t case first, referring to Fig. \ref{fig:siw_nnt}, 
where we plot the corresponding self-energy in Matsubara representation. 
In panel (a), (b), and (c) the imaginary part of the local self-energy for different values of $V/t$ is shown 
(the real part is zero due to the half-filling condition). 
One can see that the nano approximation nicely captures the local physics, accurately reproducing 
the exact self-energy at low frequencies and thus providing a reliable estimate for the quasi-particle residue 
\begin{equation} \label{eq:qpres}
 \displaystyle{Z\!=\!{\Big(1 - \frac{\partial {\rm Im} \Sigma(\omega_n)}{\partial \omega_n} \Big |_{\omega_n\!\rightarrow\!0}\Big)}^{-1}} .
\end{equation}
Moreover the amplitude of the local self-energy slightly decreases as $V$ is increased, 
the system becoming less correlated and the agreement even improving. 

However the slope of the local self-energy at $\omega\!=\!0$ in some cases is clearly not enough to capture the full picture, and we have to take the non-local self-energy into account. 
This becomes evident analysing the molecular limit $V/t\!=\!0$. 
In this limit and at finite $U$, the exact solution evidently predicts an insulating solution, 
as one can see from the absence of spectral weight at the Fermi level in the main panel of Fig. \ref{fig:A0}. 
The gap is controlled by $U$ and is due to large non-local contributions of the self-energy, 
as shown in panels (d), (e) and (f) of Fig. \ref{fig:siw_nnt}. 
At the same time, in the non-interacting limit $U\!=\!0$, the isolated, half-filled, benzene molecule is a trivial band insulator. 
In this case, the Hamiltonian is made only of the kinetic term and the gap $\Delta$ is given by the energy difference between the 
bonding and anti-bonding eigenstates, $\Delta\! \sim \!2t$. 
One can show, on the basis of simple arguments, that, in presence of non-local correlations, 
a suppression of the spectral weight at the chemical potential 
can be achieved by a large ${\rm Re}\Sigma_{i \neq j}$ even in the case of a linearly vanishing ${\rm Im}\Sigma_{ii}(\omega_n\!\rightarrow\!0)$ 
as in panel (a) \cite{valliNYP}. 
Another interesting point concerning the results of Fig. \ref{fig:siw_nnt} is that, upon increasing $V$, 
deviations from the exact results due to non-local correlations are quickly suppressed, while the local ones remain sizable. 
Approaching the limit $V\!\sim\!U$ of course also local correlations are gradually suppressed. 

These results can be summarized as follows: When the molecule is weakly connected to the contacts, one needs to go beyond DMFT, 
i.e. taking non-local correlations into account, in order to provide a good description of the system. 
On the other hand, in the region of intermediate hybridization coupling 
the most important role is played by the local physics, in a situation where the molecule is still strongly correlated. 
In many actual cases this is the interesting region from the experimental point of view. 
This suggests that our method provides an accurate tool for describing the physics of correlated nanostructures, 
already at the DMFT approximation level. When non-local spatial correlations become non-negligible, 
one needs instead to go beyond DMFT.

In addition to this, encouraging results come also from the analysis of the self-energy in the all~t configuration, shown in Fig. \ref{fig:siw_allt}. 
In panels (a) and (b) respectively, we show the comparison for the imaginary and the real part of the local self-energy at $V/t\!=\!1$, 
i.e. very close to the lowest value of $V/t$ accessible to the QMC exact solution. 
The agreement between the curves is substantially perfect as expected on the basis of the previous analysis. 
Moreover in this case, where the connectivity is higher with respect to the NN~t hopping topology, any non-local contribution to the self-energy is 
already negligible with respect to the local ones, namely, almost two order of magnitude smaller (not shown). 
This means that the region where only the local physics is important extends to lower values of 
the hybridization when the connectivity is higher. Below this threshold, as discussed before, 
we can study the evolution of the self-energy toward the molecular limit only with DMFT. 
The results, in panel (c) and (d), show that the system is becoming more correlated upon decreasing $V$, 
i.e. the imaginary part of the local self-energy is increasing, and most likely also non-local correlations arise. 
Concerning the real part of the self-energy, it displays the formation of a peak-structure at low energy, 
while the large-frequency tail, determining the filling, tends toward zero as the system gets close to half-filling 
(compare also with the inset of Fig. \ref{fig:besetzung}).

\begin{figure}[ht]
\begin{center}
\includegraphics[angle=0, width=1.00\columnwidth]{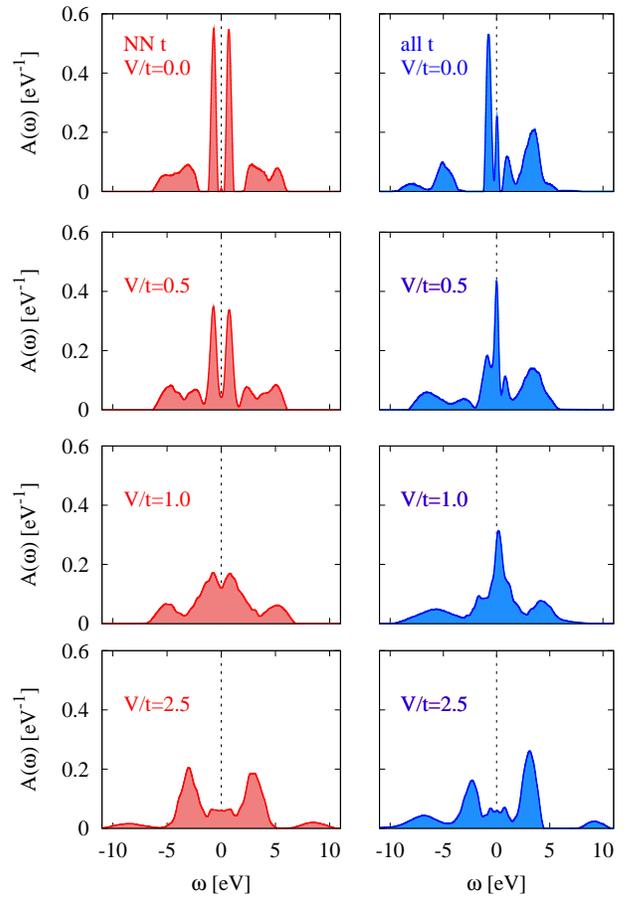}
\caption{(Color online) Evolution of the DMFT one-particle spectral function $A(\omega)$ with $V/t$, at $U\!=\!5t$ and $T\!=\!0.05t$, 
for both NN~t and all~t hopping configurations, according to the labels in the plots.}
\label{fig:bz_spectra}
\end{center}
\end{figure}

In order to have a better picture of the behavior of the system, not limited to the Fermi level, 
we present in Fig. \ref{fig:bz_spectra} the evolution with $V/t$, for both NN~t and all~t hopping configuration, 
of the spectral function $A(\omega)$, obtained via analytic continuation on the real axis 
of the DMFT(QMC) data using a Maximum Entropy method. \cite{gubernatisPRB44}
Already in the case of the isolated molecule, i.e. $V/t\!=\!0$, the spectral function of the interacting system 
shows substantial differences from the non-interacting one shown in the inset of Fig. \ref{fig:A0}. 
In the NN~t topology $A(\omega)$ retains two low energy peaks, symmetric with respect of the Fermi energy, 
corresponding to the bonding and anti-bonding peaks of the non-interacting spectrum, 
and some of the spectral weight is shifted to higher energies, forming a lower and an upper Hubbard band.
Only some spectral weight fills the band gap, which is consistent with the previous finding of $\overline{A(0)}$ (main panel of Fig. \ref{fig:A0}), 
and with a linearly vanishing DMFT self-energy at $\omega=0$ (panel (a) of Fig. \ref{fig:siw_nnt}), 
The redistribution of the spectral weight due to the interaction is instead more drastic in the all~t topology. 
The five-fold degenerate peak of the non-interacting spectrum disappears, as its degeneracy is lifted, and the resulting spectrum is metallic. 
This difference becomes even more important when the system is coupled to the leads.
The hybridization to the metallic leads, on one hand provides an additional broadening of the many-body states, via the so-called 
lead self-energy (or hybridization function), while on the other it favors the emergence of a Kondo-like resonance at the Fermi energy 
in the single particle spectral function of each of the benzene sites. 
Therefore, in the all~t topology one can observe the formation of a narrow resonant peak at the Fermi energy in the spectral function, 
while in absence of electronic correlation the structure exhibits a fairly large gap 
at the Fermi level in both hopping configurations, and no resonant peak would exist.
Upon increasing $V$ the resonance exhibits a maximum and is then suppressed. 
In the limit $V\!\gg\!t$, the hybridization becomes the dominating energy scale, 
the spectral weight is shifted to the Hubbard bands, and the spectral functions, in both hopping configurations, 
loose almost all low-energy features becoming similar to each other.

A very interesting issue deals with the study of electronic transport in correlated nanostructures. 
The conductance through the benzene ring $G(\omega)\!=\!(e^2/h) T(\omega)$ 
can indeed be computed along the lines of Refs. \onlinecite{landauerIBM1, buettikerPRL57, meirPRL68, georgesPRL82, solomonJCP129}, 
using the Meir-Wingreen generalization of the Landauer formula. 
Here $e^2/h$ is the conductance quantum, $e$ and $h$ being the electron's charge and Planck's constant respectively.  
The transmission function $T(\omega)$ is given by 
\begin{equation} \label{eq:T}
 T(\omega)= {\rm Tr}[ \Gamma G^{\rm r}(\omega) \Gamma G^{\rm a}(\omega)] ,
\end{equation}
where $G^{\rm r,a}$ are the retarded and the advanced Green's function respectively, 
and the leads scattering amplitude is given by $\Gamma=2\pi \rho V^2$. Note, that Eq. (\ref{eq:T}) for the conductance neglects vertex corrections. 
It would nevertheless be exact if {\em all} sites of the benzene molecule are coupled symmetrically to the leads 
between which the conductance is computed. \cite{meirPRL68, georgesPRL82}
However, in our case Eq. (\ref{eq:T}) without vertex correction is an approximation. 
In the case of the benzene ring, if we restrict ourselves to the Fermi level, 
we can compute the transmission function between site $i$ and $j$ from the non-local interacting Green's function we obtain from the QMC as 
\begin{equation}\label{eq:G}
 T = T_{ij}(\omega=0) = 2 \Gamma_i {|G_{ij}(\imath \omega_n \rightarrow 0)|}^2 \Gamma_j ,
\end{equation}
where the factor $2$ stems from spin degeneracy. 

\begin{figure}[t]
\begin{center}
\includegraphics[angle=0, width=0.80\columnwidth]{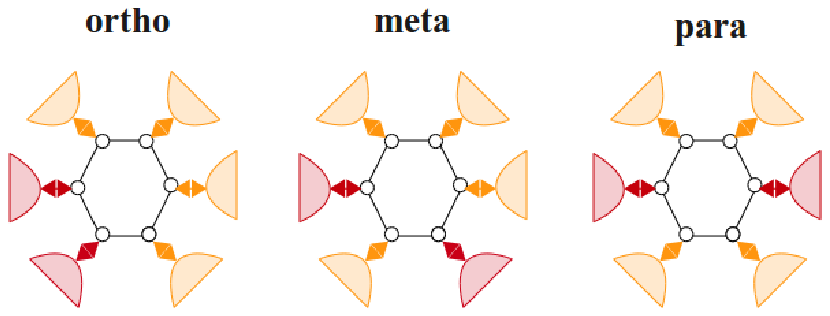}
\includegraphics[angle=0, width=0.95\columnwidth]{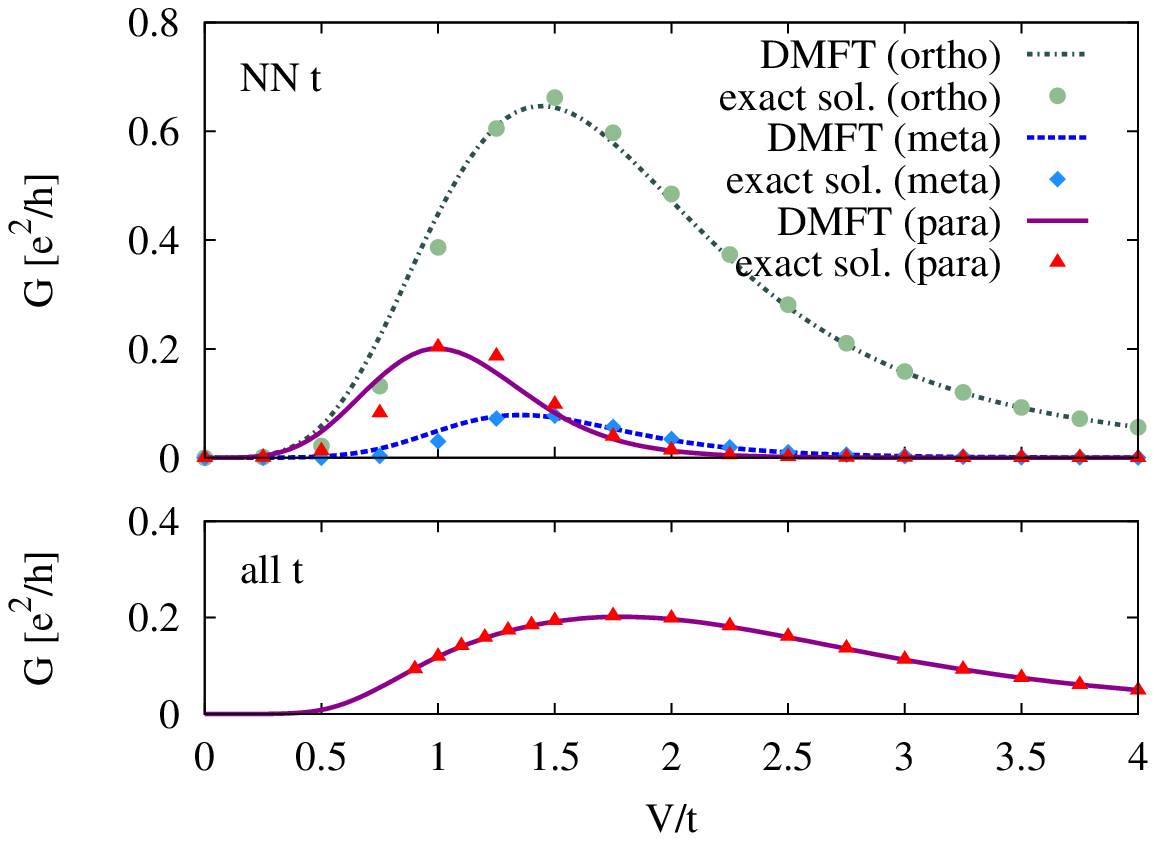}
\caption{(Color online) Conductance through the benzene ring as a function of the hybridization $V$ to the leads, 
for NN~t (upper panel) and all t (lower panel) hopping geometries at $U\!=\!5t$ and $T\!=\!0.05t$. 
In the all~t case, due to the symmetry of the problem, all connections are equivalent. 
The ortho, meta and para labels refer to the conductance computed between two of the metallic leads, as shown schematically above the plots.}
\label{fig:G}
\end{center}
\end{figure}

\begin{figure}
\begin{center}
\includegraphics[angle=0, width=0.90\columnwidth]{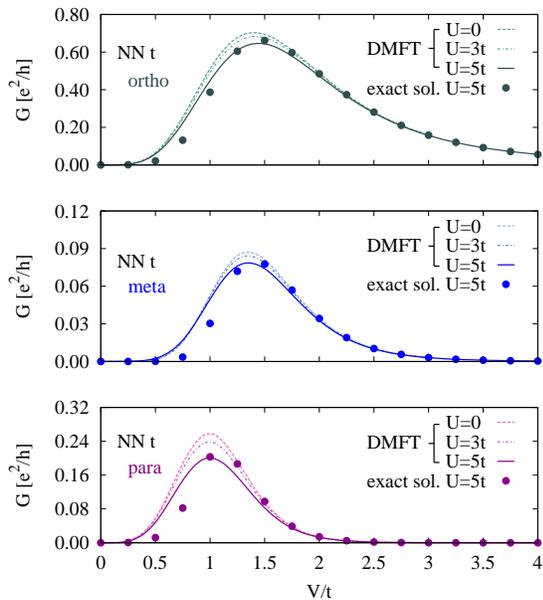}
\caption{(Color online) Conductance $G$ in the three contact positions as a function of $V/t$ 
for different values of $U$ at $T=0.05t$.}
\label{fig:GofU}
\end{center}
\end{figure}

In the literature \cite{solomonJCP129} the conductance is usually calculated in the configuration where the molecule bridges two leads only. 
Depending whether the leads are connected to the nearest, next-nearest, or next-next-nearest (i.e. opposite) neighboring-sites 
of the benzene ring, those configuration are labeled as \emph{ortho}-, \emph{meta}-, and \emph{para}-positions, respectively. 
For symmetry reasons we have instead each site of the benzene ring equivalently coupled to its own lead. 
However, with this \emph{caveat}, in the following we keep the literature nomenclature and we refer to the transmission function of Eq. (\ref{eq:G}) 
in the channel $j\!=\!i+1$, $j\!=\!i+2$, and $j\!=\!i+3$, as the ortho-, meta- , and para-position transmission through the benzene ring respectively, 
as shown in the upper panel of Fig. \ref{fig:G}. 
%This difference is not so important, i.e. the conductance profile is not substantially modified, until $\Gamma \! \lesssim \! t$.
The results for the zero-bias conductance as a function of the hybridization strength are shown in the lower panels of Fig. \ref{fig:G}.

As a general remark, valid for all connections, we can see that $G$ increases like $V^4$ at low values of the hybridization, 
as it could be expected from Eq.s (\ref{eq:G}), treating the scattering amplitudes perturbatively. 
As $V$ increases, $G$ exhibits a maximum due to the formation of a Kondo resonance between each site and its own lead, 
which is then smeared out as $1/V^2$ as a consequence of the broadening of the resonance itself. 
In the all~t topology, $G$ is the same in all three contact positions, i.e., all positions are equivalent 
due to the particular hopping structure. 
The comparison between DMFT and the exact solution shows that non-local correlations 
are not important both in the limit in which the molecule is strongly coupled to the leads and when the connectivity is high, 
which, in the light of the results presented before, may not be surprising since the conductance is computed out of 
the one-particle Green's functions, according to Eq. (\ref{eq:G}).

It is interesting to notice that, in the NN~t topology, our calculation reproduces the reduction of the conductance in the meta-position, 
with respect to the ortho- and the para-position. \cite{solomonJCP129}
This effect is believed to be a generic characteristic of single molecule junctions, 
and it has been explained in terms of quantum interference in the transmission function, 
arising only from the molecule's topology and not directly related to the presence of electronic correlations. \cite{markussenNL10}
On the other hand, many-body effects have been recently reported \cite{bergfieldNL11} to be responsible of the formation of transmission minima 
(so-called ``Mott nodes'') in molecules with open shell configurations.
It is therefore interesting to analyze the influence of $U$ on the profile of the zero-bias conductance.
In Fig. \ref{fig:GofU} we report the results of our calculations at different values of $U/t$ in the whole hybridization range. 
Note that in the non-interacting limit also DMFT is obviously exact. 
The main effect of $U$ is to suppress the conductance peak, while the low- and high- hybridization regimes are not much affected. 
We compare the {\it percentage-wise} reduction $\Delta(U)$ of the conductance maximum at $U\!\neq\!0$ 
with respect to its non-interacting value (at the same value of $V$) 
in order to get information on the effect of correlations \emph{on top} of the topological reduction.
%, according to the formula 
%\begin{equation} \label{Eq:Delta}
% \displaystyle{\Delta(U)\!=\!\frac{{G}^{\textrm{max}}(0)-
%G^{\textrm{max}}(U)}{G^{\textrm{max}}(0)}}. 
% \displaystyle{\Delta_{\textrm{label}}(U)\!=\!\frac{G_{\textrm{label}}^{\textrm{max}}(U\!=\!0)-
%G_{\textrm{label}}^{\textrm{max}}(U)}{G_{\textrm{label}}^{\textrm{max}}(U\!=\!0)}} , \label{Eq:Delta}
%\end{equation}
We find out that the suppression increases with the distance between the sites through which the conductance is computed, i.e. 
$\Delta_{\textrm{ortho}}<\Delta_{\textrm{meta}}<\Delta_{\textrm{para}}$, 
as the Hubbard repulsion tends to localize the electrons in the molecule. 
We summarize the corresponding values in Table \ref{tab:delta}. 

\begin{table}[]
\begin{center}
 \begin{tabular}{ c c c c c c }
  \hline
  \hline \\[-7pt]
  & \textcolor{white}{space} & \textcolor{white}{space} & $\Delta(U)$ & \textcolor{white}{space} & \textcolor{white}{space} \\[1pt]
  & \textcolor{white}{morespacehere} & \textcolor{white}{PS} ${\textrm{ortho-}}$ & \textcolor{white}{PS} ${\textrm{meta-}}$ & \textcolor{white}{PS} ${\textrm{para-}}$ &  position \\[5pt]
  \hline \\[-3pt]
  & \textcolor{white}{space} $U\!=\!5t$ & $\sim 7\%$ & $\sim 9\%$ & $\sim 22\%$ & \textcolor{white}{space}\\[5pt]
  & \textcolor{white}{space} $U\!=\!3t$ & $\sim 2\%$ & $\sim 3\%$ & $\sim 7\%$ & \textcolor{white}{space}\\[5pt]
  \hline
  \hline
 \end{tabular} 
\end{center}
\caption{Percentage-wise reduction of the maximal conductance $\Delta(U)$ for ortho, meta, and para connections.}
\label{tab:delta}
\end{table}

\subsection{Quantum junctions}\label{Sec:QPC}
In the current state-of-the-art of nanoscopic electronic transport, quantum junctions play a fundamental role. 
They can be experimentally realized by a mechanically-controlled break (MCB) process of a 
metallic wire made e.g., of Au, resulting in atomically sharp contacts with an adjustable tunneling gap. 
Strong evidences from conductance quantization have been reported, both at low \cite{vanweesPRL60} and at 
room temperature, \cite{mullerPRB53} as a proof of the experimental realization of single atomic junctions. 
Moreover molecules can be adsorbed into the gap, forming stable tunneling contacts, and
allowing for the observation of electronic transport through molecular systems. \cite{reedSci278}

One can expect electronic correlations to become relevant in the contact region, where electrons are spatially confined in narrow structures, 
as well as in the bridging molecule itself.
In Ref. \onlinecite{valliPRL104} we carried out a calculation on a model for a quantum junction 
made out of more than hundred correlated sites, showing that our method is able to handle even very complex nanostructures.

\begin{figure}[hb!]
\begin{center}
\includegraphics[angle=0, width=0.8\columnwidth]{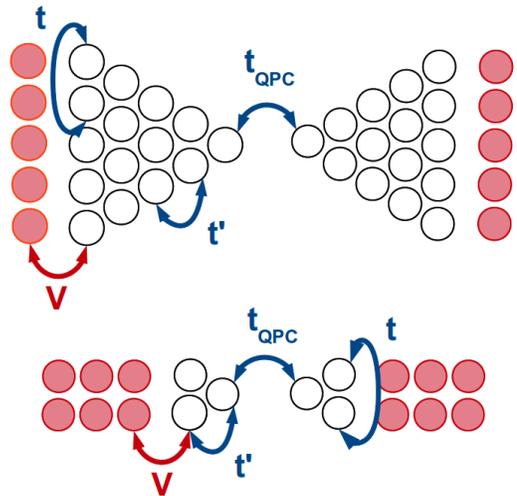}
\caption{(Color online) Scheme of the (3-dimensional) quantum junctions with five (upper panel), 
and two (lower panel) layers. 
Empty atoms have an on-site Hubbard repulsion $U$, and are connected via hopping $t$, $t'$ and hybridization $V$ channels 
as schematically shown in the pictures.}
\label{fig:toyqj}
\end{center}
\end{figure}

\begin{figure*}[t!]
\begin{center}
\includegraphics[angle=0, width=2.00\columnwidth]{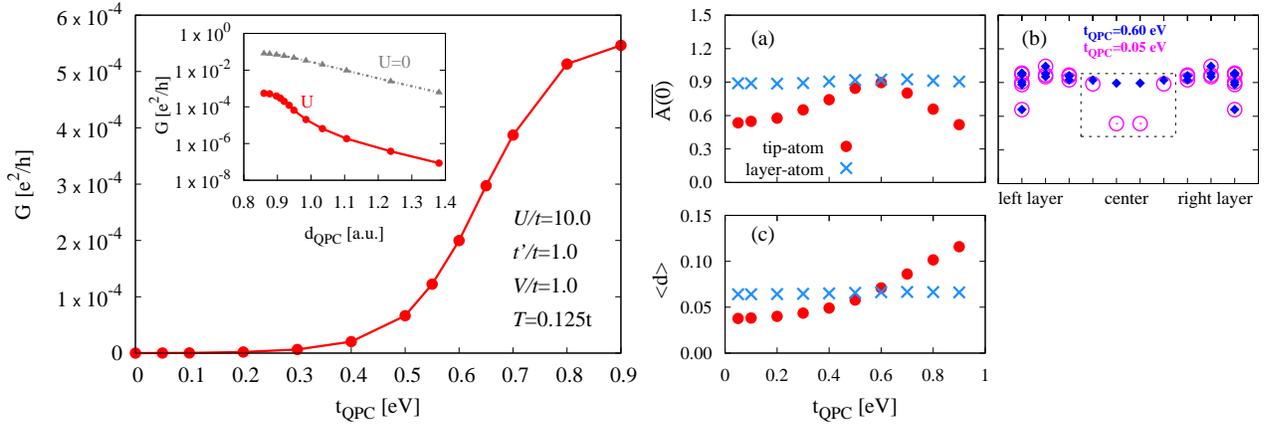}
\caption{(Color online) Left panel: Conductance $G$ through the five-layer QPC as a function of $t_{\rm QPC}$. The parameters are shown in the plot. 
Inset: G as a function of an estimate of the inter-tip distance $d_{\rm QMC}$, as explained in the text.  
The data of the main panel (circles) are compared to the non-interacting case (triangles) on a logarithmic scale 
in order to highlight the effect of electronic correlations. 
Right panels: Atom-resolved low-energy spectrum $\overline{A(0)}$ double occupations $\langle d \rangle$. 
In panels (a) and (c) the dependence on $t_{\rm QPC}$ for the low-energy spectrum and the double occupation is shown 
for the tip-atom and an atom sitting in the layer closest to the tip. 
In panel (b) we show the low-energy spectrum for each inequivalent atom of the structure for two values of $t_{\rm QPC}$, 
namely $t_{\rm QPC}\!=\!0.60$~eV (diamonds), and $t_{\rm QPC}\!=\!0.05$~eV (dotted circles). The change of $t_{\rm QPC}$ only affects the tip-atoms. 
The region labeled with ``center'' and highlighted in the plot corresponds to the atoms analyzed in panels (a) and (c).}
\label{fig:qpc}
\end{center}
\end{figure*}

Before discussing the physical results we obtained, it is useful to recall the characteristic of the model junction. 
The system is made of two identical structures of correlated atoms with a simple body-centered cubic (bcc) lattice symmetry, 
narrowed in a double-cone-like junction. 
For the sake of simplicity, we assume a single band model which may be suitable for cuprate or cobaltate junctions, 
though orbital selective tunneling processes will probably also play a role in, e.g. gold or aluminum devices.
Hopping processes are allowed to nearest neighbor sites, both intra-layer and inter-layer, with amplitudes $t$ and $t'$ respectively, 
while the hopping between the tips is defined by the parameter $t_{{\rm QPC}}$. 
The outermost layer of each structure is connected via hybridization channels $V$ to non-interacting leads, describing e.g. the bulk-like atoms of the wire. 
A scheme of junctions of different sizes we will discuss below is shown in Fig. \ref{fig:toyqj}.

In the previous Letter \cite{valliPRL104} we addressed the problem of what is happening at the junction in the MCB process, and 
we simulated the breaking of the junction by changing the distance $d_{\rm QPC}$ between the two structures 
thought the control parameter $t_{{\rm QPC}}$, i.e. the overlap of the electrons' atomic-like wave functions of the tip-atoms. 
According to Ref. \onlinecite{blancoPSS81}, $t_{\rm QPC}\!\sim\!(1/d_{\rm QPC}^{l+l'+1}){\rm exp}(-d_{\rm QPC})$, 
where $l$ and $l'$ are the angular momentum quantum numbers associated to the orbitals involved in the tunneling process. 
In the following we calculate the distance $d_{QPC}$ according to the above formula, 
where we suppose $l\!=\!l'\!=\!2$, i.e. a $d$-like orbital character for the correlated atoms bands, 
however the general physical argument discussed in the following does not depend on the precise dependence of $d_{\rm QPC}$ on $t_{\rm QCP}$. 

We calculate the total transmission function through the junction summing over all possible transmission channels 
\begin{equation}
 T =  \sum_{ij} T_{ij} = 2 \frac{e^2}{h} \sum_{i\in{\rm L}} \sum_{j\in{\rm R}} \Gamma_i {|G_{ij}(\imath \omega_n \rightarrow 0)|}^2 \Gamma_j ,
\end{equation}
where ${\rm L}$ (left) and ${\rm R}$ (right) correspond to the correlated atoms sitting in the outermost layers of the junction. 
As $d_{\rm QPC}$ is increased DMFT reveals strong deviations from an exponential behavior of $G$, 
expected in the case of a tunneling process through a barrier. Such a result is associated to a local Mott-Hubbard crossover, occurring at the tip-atom(s). 
It can be explained considering that the MCB process effectively removes a neighbor from the already poorly connected tip-atoms. 
This further reduces the metallic screening of the local Coulomb interaction expected in the bulk, causing the tip to become more insulating-like. 
If correlations can so strongly influence the electronic structure of the contacts, 
this phenomenon could have a huge impact on the interpretation of experimental results, and it is therefore worth being investigated extensively.

The fundamental question to answer is whether the crossover is a generic characteristic of quantum junctions.
Following Ref. \onlinecite{valliPRL104}, we performed calculations of the 5-layer QPC (shown in the upper panel of Fig. \ref{fig:toyqj}) 
but for a slightly different set of parameters, more realistic for the bcc geometry considered here, 
where the intra- and inter-layer hopping amplitudes $t$ and $t'$ are the same, and we choose $t\!=\!0.40$~eV. 
The results for such a structure are shown in Fig. \ref{fig:qpc}. 
In the left panel we show the conductance $G$ through the five-layer QPC, as a function of $t_{{\rm QPC}}$, 
while we reproduce the same data as a function of the inter-tip distance $d_{\rm QPC}$ comparing to the non-interacting case on a logarithmic scale, 
so that the fingerprint of the Mott-Hubbard crossover, and its effect on electronic transport is highlighted: 
$G$ shows a more-than-exponential suppression as a function of $d_{\rm QPC}$ for a value of $d_{\rm QPC}$ 
corresponding to $t_{{\rm QPC}}\!\sim\!t$, and recovers instead an exponential behavior at larger distances. 
In order to show that the relevant physics concerns the tip-atoms, one can consider atom-resolved local quantities, 
i.e. the occupations and the local low-energy spectrum.  
We find that, due to the strong electronic correlations (provided that $U/t$ is large enough, as it will be clarified in the following) 
the occupation $\langle n \rangle$ of all the atoms of the QPC stay at half-filling in the whole range of $t_{\rm QPC}$. 
Moreover, also the double occupations $\langle d \rangle$ and the low energy spectrum $\overline{A(0)}$ 
of all layer atoms (i.e., all atoms except for the tip ones) are almost constant in the whole $t_{\rm QPC}$ range, 
their values naturally depending on the position of the atoms in the junction. 

\begin{figure}[t!]
\begin{center}
\includegraphics[angle=0, width=1.00\columnwidth]{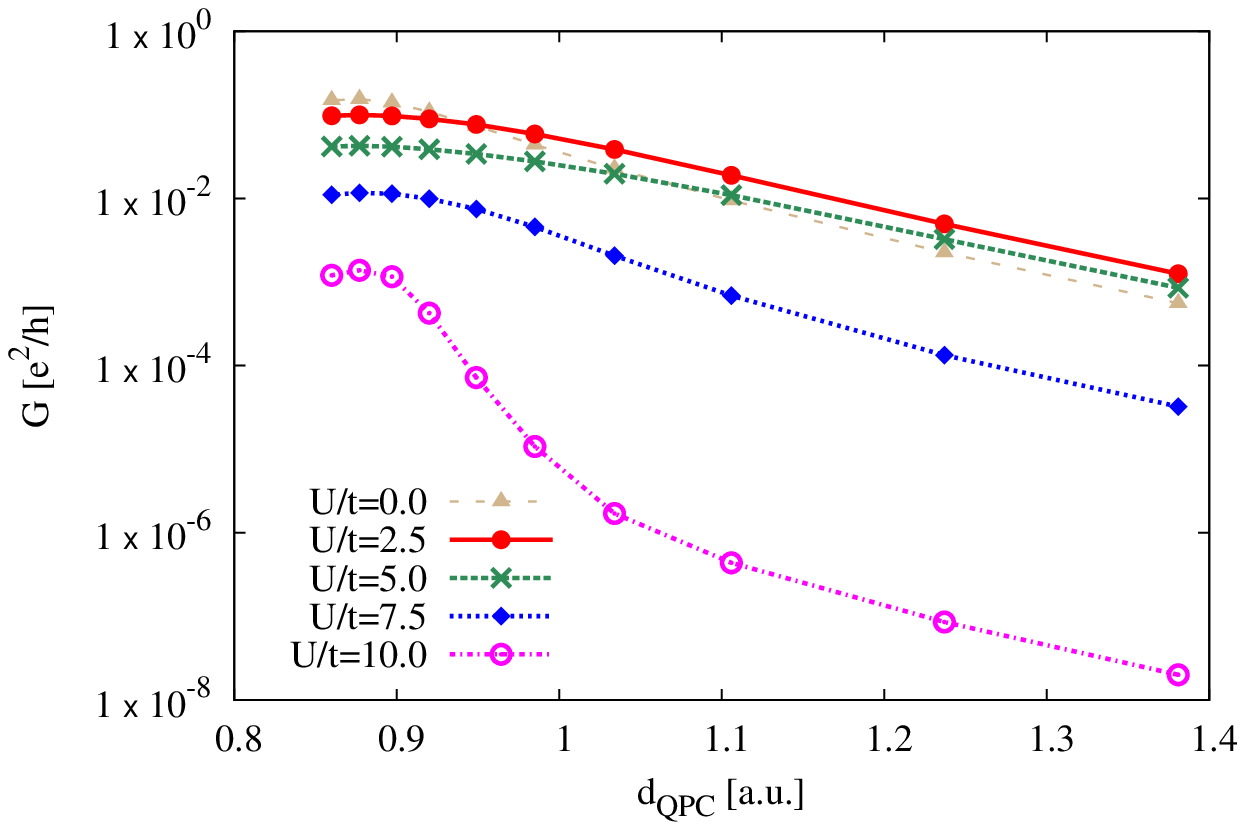}\\[-0.3cm]
\includegraphics[angle=0, width=1.00\columnwidth]{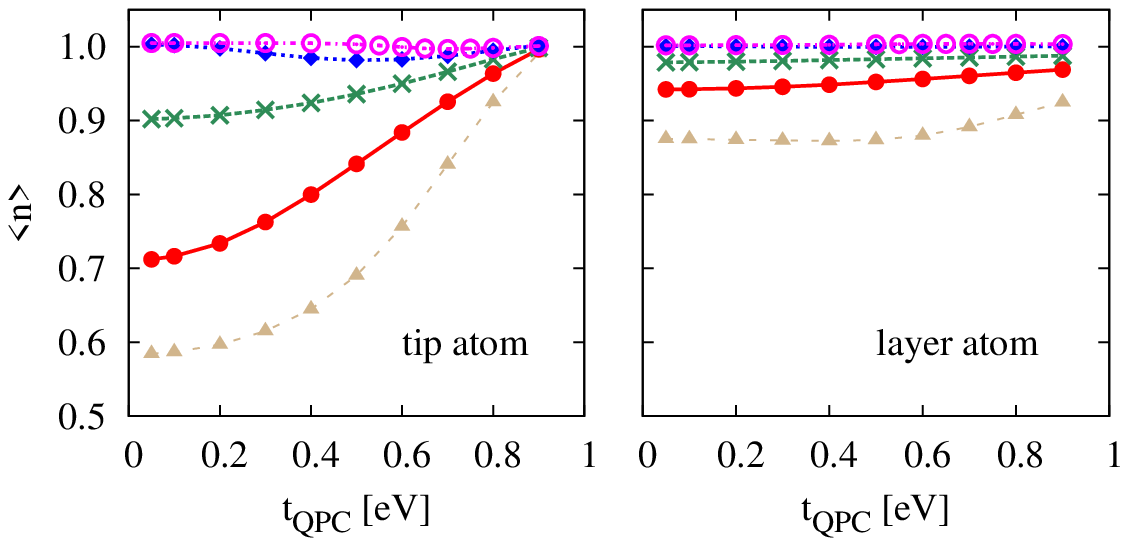}\\[-0.3cm]
\caption{(Color online) Upper panel: conductance $G$ through the two-layer QPC vs. $d_{{\rm QPC}}$ for different values of $U/t$. 
Lower panel: corresponding density of the two inequivalent atoms according to the labels in the plot.}
\label{fig:q2lj}
\end{center}
\end{figure}

\begin{figure}[ht]
\begin{center}
\includegraphics[angle=0, width=1.0\columnwidth]{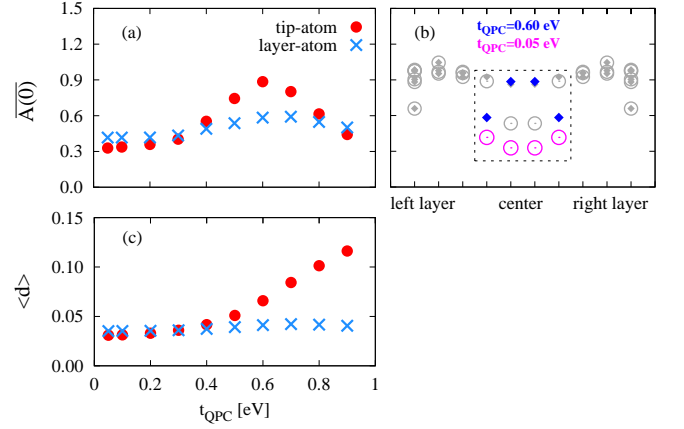}
\caption{(Color online) Left panels: atom-resolved double occupations $\langle d \rangle$ and low-energy spectrum $\overline{A(0)}$ as a function of $t_{\rm QPC}$ 
at $U\!=\!10t\!>\!U^*$ and $T\!=\!0.125t$. 
Right panel: $\overline{A(0)}$ for each of the structure inequivalent atoms for two values of $t_{\rm QPC}$, 
namely $t_{\rm QPC}\!=\!0.60$~eV (diamonds), and $t_{\rm QPC}\!=\!0.05$~eV (dotted circles). 
The region labeled with ``center'' and highlighted in the plot corresponds to the atoms analyzed in panels (a) and (c). 
Light shaded data are the corresponding values of the spectrum of the five-layer QPC, for comparison.}
\label{fig:q2lj2}
\end{center}
\end{figure}

\begin{figure}
\begin{center}
\includegraphics[angle=0, width=1.00\columnwidth]{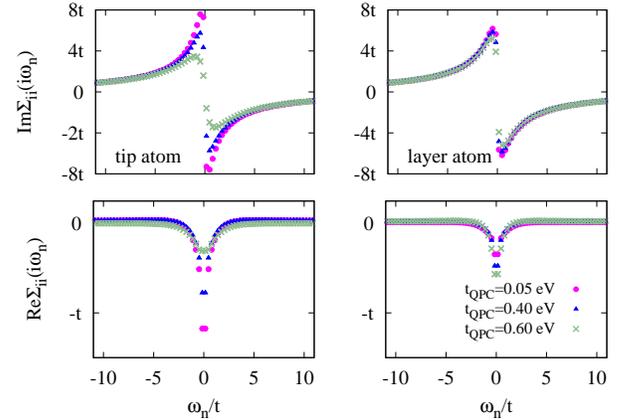}
\caption{(Color online) Layer-resolved DMFT local self-energy in Matsubara representation of the only two inequivalent atoms 
(i.e., tip-atom and layer-atom, according to the labels in the plot) of the two-layer QPC 
for $U\!=\!10t\!>\!U^*$ and $T\!=\!0.125t$ for different values of $t_{\rm QPC}$. 
While the tip-atom is strongly affected by $t_{\rm QPC}$ the layer-atom exhibits no significative dependence.}
\label{fig:q2ljSiw}
\end{center}
\end{figure}

We find that a completely different behavior characterizes the tip-atoms instead. 
%According to the argumentation above,
In the right panels of Fig. \ref{fig:qpc} we restrict ourselves, for convenience, 
to the comparison of local quantities of just two representative atoms in the QPC, namely the tip-atom and one of the atoms 
sitting in the layer directly connected to the tip (denoted as layer-atom in the following). 
Below $t_{\rm QPC}\!=\!0.60$~eV, both $\langle d \rangle$ and $\overline{A(0)}$ are continuously and monotonically suppressed as $t_{\rm QPC}$ decreases, 
confirming that the tip-atoms undergoes a local Mott crossover. 
One may further notice that the tips' $\overline{A(0)}$ reaches a maximum around $t_{\rm QPC}\!=\!0.60$~eV 
and decreases again for values of $t_{\rm QPC}$ above this threshold. 
This phenomenon is not correlation-driven, as the double occupations always increase with increasing $t_{\rm QPC}$, 
but is caused by the recombination of the tip-atoms' states into a bonding and anti-bonding structure.

In order to better understand the physics behind this phenomenon, we significantly reduce the complexity of the problem. 
We consider a junction made of two layers (shown in the lower panel of Fig. \ref{fig:toyqj}) 
so that the tip-atom is not connected directly to a bath of free electrons, but to a layer of correlated atoms. 
Therefore, there are by symmetry only two inequivalent atoms left, tip- and layer-atoms. 
This makes the system much simpler, but still allows us to observe a dichotomy between the two kinds of correlated atoms. 
Nevertheless it is important to stress that even this minimal model for the quantum junction cannot be solved exactly with QMC 
due to a severe fermionic sign problem.

From the evolution of the conductance through the junction for different values of $U$, it is clear 
that some critical value $U^*$ exists, above which $G$ exhibits the more-than-exponential behavior associated with the Mott-Hubbard crossover, 
as shown in Fig. \ref{fig:q2lj}. 
The non-interacting QPC shows no peculiar feature, it evolves smoothly 
from the contact (small $d_{\rm QPC}$) to the tunneling (large $d_{\rm QPC}$) regime. 
As the value of $U$ is increased the conductance is globally suppressed, and above some threshold it develops 
a much faster transition between the contact and tunneling regimes. 
In all cases $G$ does not reach the limit $G_0\!=\!e^2/h$ in the contact regime because of the absence of a completely open transmission channel.
Similar observation on MCB junctions \cite{kransPRB48} or on the tunneling spectra of Co impurities adsorbed on a Cu(100) surface \cite{néelPRL98} 
support the hypothesis that such effects, which we show to be driven by correlations, can be of importance in experiments.
In order to relate the behavior of $G$ with a Mott-Hubbard transition, we also look at the layer-resolved $\overline{A(0)}$, 
in analogy to the five-layer QPC. One has anyway to be careful here, 
because below a critical $U$, the spectral weight is affected by strong, $t_{{\rm QPC}}$-dependent, density fluctuations. 
On the other hand, approaching $U^*$ all atoms occupations tend towards half-filling (see Fig. \ref{fig:q2lj} lower panels) 
and the analysis of the spectral weight for this porpoise is safe. 

In Fig. \ref{fig:q2lj2} we show the results for the low-energy spectrum $\overline{A(0)}$ and the double occupations $\langle d \rangle$ 
at a value of $U$ above the critical one ($U>U^{*}$), analogously to the case of the five-layer QPC. 
We can observe that below the threshold value $t_{\rm QCP}\!=\!0.60$~eV both $\langle d \rangle$ and $\overline{A(0)}$ 
for the tip-atoms decrease continuously with $t_{\rm QPC}$ as the tip-atom becomes more insulating-like. 
The overall behavior for the tip-atom of the different size QPCs is qualitatively very similar, 
suggesting the existence of a (size-independent) energy scale associated with this phenomenon. 
One can notice that in the two-layer QPC the layer-atoms' local quantities show some dependence on $t_{\rm QPC}$,  
and the values of the double occupations and of the low-energy spectrum of the layer-atoms are slightly reduced 
compared to the corresponding one of the five-layer QPC. 
This indicates correlation effects to be further enhanced upon decreasing the system size. 

One can indeed check that, considering junctions of intermediate size with respect to the ones shown here, 
i.e. adding one layer after the other to the two-layer QPC, helps the stabilization of both the local quantities of the layer-atoms, 
which loose almost any dependence on $t_{\rm QPC}$, while the tip-atoms still display a local Mott-Hubbard transition.

Another clear evidence of the enhancement of electronic correlations in the MCB process 
is also provided by the evolution with $t_{\rm QPC}$ of the DMFT self-energy. 
In Fig. \ref{fig:q2ljSiw} we show the imaginary and real part of the local self-energy for both inequivalent atoms of the two-layer QPC 
at $U\!=\!10t$ ($U>U^{*}$).
As already mentioned, the MCB process confines the tip-atom(s) at the edge of the structure, and drastically suppressing the hopping channel in one direction 
further reduces the screening of the Coulomb repulsion. Consequently, electronic correlations in the tip-atom are strongly enhanced 
determining a local Mott-Hubbard crossover in the tip, while the layer-atom self-energy does not show significant dependence on $t_{\rm QPC}$. 
Below the critical $U$ instead, one finds a dependence on $t_{\rm QPC}$ only in ${\rm Re}\Sigma_{ii}(\imath \omega)$, 
i.e., in the renormalization of the chemical potential, responsible for the change in the occupancy (not shown). 
The above analysis shows once more that the change in the conductance in the MCB process can be traced back to  
the strong electronic correlations arising from the spatial confinement of electrons in sharp contact devices. 

\section{Conclusions and Outlook}\label{Sec:conclusion}
We have studied electronic correlations in nanoscopic systems 
within a many-body approach suitable to deal with complex correlated structures. 
We show that including local electronic correlations we can reasonably describe nanoscopic systems 
with many neighbors, long range hopping, or a sufficiently strong hybridization to non-interacting leads. 
These conditions are fulfilled in many cases of interest, but there are regimes in which non-local self-energies are observed 
and it becomes necessary to also include spatial correlations beyond DMFT. 
We therefore plan to generalize the present method within the framework of D$\Gamma$A, 
in order to include spatial correlations at all lenght scales. This way we expect to be able to recover a reliable description 
of the nanoscopic system on an even larger parameter range. 
The present approach can be viewed as the $n\!=\!1$ particle level of a more general nanoscopic D$\Gamma$A scheme. 
It could be an important tool for investigating electronic transport in correlated structures at the nanoscale. 
In this respect we also studied, as a potential application, the transport through a MCB junction. 
Going beyond Ref. \onlinecite{valliPRL104}, we investigated the phenomenon of a local Mott-Hubbard crossover 
in quantum junctions of different sizes, showing that it is a general feature of sharp nanostructures 
and that it may be of importance for the interpretation of experiments.

\section*{}
\begin{acknowledgments}
We would like to thank in particular Olle Gunnarsson who strongly contributed to the early stages of this study 
and S.~Andergassen for suggestions and a very careful reading of the manuscript. 
Discussions with M.~Capone, C.~Castellani, S.~Ciuchi, C.~Di~Castro, J.~Lorenzana, G.~Rohringer, and D.~Rotter are also acknowledged. 
We acknowledge financial support from the Austrian Science Fund (FWF) through 
F4103-N13 (AV), I610-N16 (AT), and I597-N16 (KH). 
Calculations have been performed on the Vienna Scientific Cluster. 
\end{acknowledgments}


\begin{thebibliography}{99}
\bibitem{footnote1} A realistic calculation of Benzene might also require to include non-local Coulomb interactions. \cite{linPRL106} 
This is only possible in a suggested extension of D$\Gamma$A. \cite{toschiAP523}
\end{thebibliography}
\end{document}